# Experience with Distributed Memory Delaunay-based Image-to-Mesh Conversion Implementation

Polykarpos Thomadakis and Nikos Chrisochoides[1]
Center For Real-Time Computing
Old Dominion University

**Abstract**

This paper presents some of our findings on the scalability of parallel 3D mesh generation on distributed memory machines. The primary objective of this study was to evaluate a distributed memory approach for implementing a 3D parallel Delaunay-based algorithm that converts images to meshes by leveraging an efficient shared memory implementation. The secondary objective was to evaluate the effectiveness of labor (i.e., reduce development time) while introducing minimal overheads to maintain the parallel efficiency of the end-product i.e., distributed implementation.

The distributed algorithm utilizes two existing and independently developed parallel Delaunay-based methods: (1) a fine-grained method that employs multi-threading and speculative execution on shared memory nodes and (2) a loosely-coupled Delaunay-refinement framework for multi-node platforms. The shared memory implementation uses a FIFO work-sharing scheme for thread scheduling, while the distributed memory implementation utilizes the MPI and the Master-Worker (MW) model.

The findings from the specific MPI-MW implementation we tested suggest that the execution on (1) 40 cores not necessary in the same single node is 2.3 times faster than the execution on ten cores, (2) the best speedup is 5.4 with 180 cores again the comparison is with the best performance on ten cores. A closer look at the performance of distributed memory and shared memory implementation executing on a single node (40 cores) suggest that the overheads introduced in the MPI-MW implementation are high and render the MPI-MW implementation 4 times slower than the shared memory code using the same number of cores . These findings raise several questions on the potential scalability of a "black box" approach, i.e., re-using a code designed to execute efficiently on shared memory machines without considering its potential use in a distributed memory setting.

## 1. Introduction

Mesh generation and refinement software are crucial for function approximation with Finite Element/Volume Methods (FEM) [1, 2]. Sequential mesh generation was used in the mid-80s (and, in some cases, still is) to set up and initialize the data structures for parallel large-scale approximation codes. The setup phase may cause a significant slowdown in the overall performance of FEM software. For instance, a study [3] found that the cost of initializing distributed memory parallel iterative linear solvers' data structures was about 30 times higher (mainly due to the lack of parallel I/O libraries) than the cost of performing 100 iterations of solving a specific (Elliptic) Partial Differential Equation (PDE) using a Semi-Iterative linear solver [4].

The issue persisted in the mid-90s despite the availability of effective parallel I/O libraries, and we suggested parallel mesh generation as a solution to enhance the performance of the setup phase for parallel large-scale field solvers [5,6]. Although the solution solved the I/O problems, the setup phase was difficult due to increased software complexity for parallel mesh generation. Developing industrial strength mesh generation software is a labor-intensive process that takes around 100 person-years. Regarding parallel mesh generation, the software complexity increases by an order of magnitude due to managing network and memory hierarchies, concurrency, workload balancing due to adaptive mesh refinement, and support for power optimizations and heterogeneous architectures.

---
[1] Corresponding author, email: nikos@cs.odu.edu

In the early 2000s, we tackled the issue of software complexity in parallel mesh generation codes by separating algorithm correctness from performance portability. This approach is even more crucial now with highly heterogeneous High-Performance Computing (HPC) parallel platforms. For over 20 years, we've developed a runtime system [7, 8, 9, 10, 11] that uses domain-specific abstractions to separate parallel architecture intricacies from algorithms. This simplifies the parallel algorithm and implementation and improves the efficiency of parallel mesh generation and refinement. Because of this "separation of concerns", we were also able to tackle more complex practical aspects, such as transitioning from 2D to 3D geometries, experimenting with different meshing techniques, managing real-time 2D/3D and even 4D sensor data like medical images, interacting with CAD systems and solvers, and incorporating metric-based anisotropy for local reconnection methods.

We tried to implement the newest parallel image-to-mesh conversion algorithm based on distributed memory Delaunay method (presented in [12]) on a parallel runtime system [10, 11]. We aimed to evaluate the impact of trade-offs due to the "separation of concerns," i.e., measure the impact of our approach on software complexity and end-user productivity while evaluating the actual performance of the end-to-end code, i.e., runtime system plus an application. We will report our findings elsewhere, as it is beyond the scope of this paper. However, in the process, we stumbled into an interesting issue while analyzing the performance of the earlier hand-coded MPI + Threads implementation of the parallel mesh generation algorithm and its implementation presented in [12]. In this paper, we present new findings (based on this experience) that call for reevaluating our previous notions about the ability to scale parallel mesh generation in the exascale era of supercomputers.

We tested extensively the MPI-MW implementation using up to 200 cores for fixed-size problems with approximately 47 million elements. The best performance was achieved for 180 cores (90 MPI ranks with two cores per rank) and it was 5.4 times faster than using 10 cores (10 MPI ranks with one core per rank). Although the speed-up was not linear, we consider this result noteworthy. However, it is important to note that the MPI-MW execution on 180 cores was more than twice slower than the execution of the shared memory multi-threaded code using 40 cores on a single node. While the execution of the shared memory on 40 cores (in a single node) is seven times faster than the execution of the MPI-MW on 40 cores using multiple nodes. This suggests that the overheads introduced in the multi-threaded code to scale for more than one node are quite high (more than 50% of the actual refinement time) and reduce its efficiency to about 50% for distributed memory platforms.

Although these findings are specific to the MPI-MW implementation we tasted, they raise several questions about the expectations for achieving scalability using this study's "black-box" approach.

## 2. Background

Parallel mesh generation methods decompose the original mesh generation problem into smaller subproblems that are solved (i.e., meshed) concurrently [13]. The subproblems can be either tightly-coupled [14,15], partially-coupled [16,17], weakly-coupled [18,19], or completely decoupled [20,21]. The coupling of the subproblems determines the intensity of the communication and the amount/type of synchronization required from the algorithm used in each subproblem. In this paper, each subproblem (or subdomain) uses the Delaunay mesh generation algorithm. The method follows a two-layered approach: (1) a tightly-coupled method presented in [15] and is designed to run efficiently at the cc-NUMA multi-core single-node, and (2) a partially-coupled method presented in [22], works correctly and efficiently for uniform mesh refinement [23] and is designed to run efficiently at the distributed memory multi-node platforms.

In [15], we presented a tightly coupled method that relied on optimistic (or speculative execution) to explore concurrency for Delaunay-based methods at the cavity[2] level. This paper briefly refers to it as PODM (Parallel Optimistic Delaunay Method). PODM is set to achieve the following objective: generate a high-quality surface and volume mesh by accurately representing the (segmented) surface of 3-dimensional (3D) medical images (i.e., biological objects). PODM's implementation on shared memory cc-NUMA machines relies on (1) a task execution model [24, 25] where the tasks share the data and use low-level locking mechanisms and contention managers and (2) load balancing schemes using multi-threaded implementations [26, 27] to enhance parallel performance with minimal overhead. An evaluation was conducted on the Pittsburgh Supercomputing Center's Blacklight, a Distributed Shared Memory (DSM) machine. The results showed that it had good *strong and weak scaling efficiency of about 80% for up to 64 and 144 cores,* respectively. However, for a larger number of cores, PODM suffers from communication overheads caused by many remote memory accesses (at the page level partly because of false-sharing due to the idiosyncratic way the points/elements are created and inserted by the Delaunay method). As a result, PODM's performance deteriorates for a core count beyond 144 due to network congestion [15].

To further optimize PODM's communications overheads for many more cores, in [28], we attempted to "constrain" the Delaunay method at each core to refine elements allocated in the same or nearby pages by using data decomposition (i.e., explore memory affinity) as opposed to domain decomposition (i.e., explore geometric affinity). Performance data indicated good performance for up to 200 cores for cc-NUMA machines. In our quest to further improve the effective use of a higher number of cores, in [29], we considered again DSM cc-NUMA machines, but with domain decomposition relying on Parallel Delaunay Refinement (PDR) method initially presented in [29], but its distributed memory implementation first presented in [12] using MPI –this paper and data presented here improve the earlier PDR MPI-MW implementation in terms of several communication aspects (see Section 3.2). Throughout this paper, we refer to this implementation as the PDR.PODM method. PDR was chosen because it uses an efficient parallel point-insertion technique and guarantees the correctness of the parallel algorithm (i.e., Delaunay property is maintained everywhere as opposed to other methods distributed parallel Delaunay mesh generation methods published in the literature). Moreover, given that PODM [15] and PDR preserve internal boundaries for multi-material (tissues) [22, 31], the PDR.PODM preserves them, too. The multi-tissue mesh generation is a very interesting and practical problem with many (mostly topological) challenges, and it will not be addressed here; however, the reader could look into earlier work (and cited publications) in [32, 33, 34] in addition to work we presented in [15].

---

[2] The cavity represents the set of elements, edges, and vertices affected by inserting a new point. These elements must be modified or refined to maintain the Delaunay property, ensuring that no point lies within the circumsphere of any triangle or tetrahedron in the mesh.

In constant time, the PDR method analyzes the dependencies between the concurrently inserted points and determines if they can be inserted independently. This approach reduces network congestion by eliminating runtime checks for data dependencies in distant regions of space- and memory-wise. PDR uses a spatial decomposition tree to split the list of candidate points (to be inserted) into smaller (local) lists that can be processed concurrently. The method starts by constructing a *coarse background* mesh for the parallel refinement step. A trade-off exists between the degree of concurrency one could exploit and the overhead for generating the background mesh. A potential shortcoming is the method's requirement to have access to sub-meshes of two layers of adjacent subdomains, i.e., excessive data dependencies and, thus, data movement. This leads to large data movements and in the presence of data-intensive (as opposed to compute-intensive) like Delaunay meshing that can underline the scalability of the method. In addition, the coarse background mesh must be refined enough to ensure sufficient (and correct) concurrency for parallel refinement. However, constructing such a dense mesh prolongs the low-concurrency part of the computation. Initially, PODM is used to generate the coarse background mesh. PODM concurrently refines each subdomain utilizing a group of cores from the DSM cc-NUMA machine within the limits of the coarse decomposition dependencies.

In [12], we tried to adapt the PDR.PODM method for distributed memory machines. This was because fewer supercomputing centers were using large DSM cc-NUMA machines, and distributed memory machines could access more ( > 200) cores. We used Massage Passing Interface and Master/Worker (MPI-MW) execution model at the coarse grain level presented in [12]. However, a clarification is required: *the worker process runs PODM with N-cores available to any given MPI rank; in each MPI rank, PODM [15] implements the FIFO work-sharing model, which is used in this study, too.* So, the Master (at the MPI rank level) could be thought of as the initialization thread that receives the subdomain to be refined, does the packing/unpacking of data, sets up the data for PODM, and then the rest of the PODM threads manage the workload as it is described in [15]. In [12], the hope was to optimize communication by explicitly managing data movement instead of large cc-NUMA DSM machines where the Operating System transparently manages page movement. The data reported in [12] indicate about 50% weak scalable efficiency for about 900 cores. We could not reproduce all the data in [12].

In our recent effort to port the MPI-MW code on the runtime system [11] for exascale machines, we stumbled upon a new data set that raised several interesting questions regarding the scalability of the specific MPI-MW-based PDR.PODM method/implementation and parallel mesh generation in general. Before we continue with the presentation of the performance data, it is worth refreshing some fundamental concepts (understanding) of parallel mesh generation (and computing more general) by considering the following remarks:

- The number of MPI ranks determines the maximum concurrency at the system's coarse-grain (subdomain) level; of course, with PODM, we could increase this concurrency by multiplying it with the concurrency at the medium-grain (cavity) level at some cost due to thread interference between communication and refinement modules. It's important to understand that even if the workload is evenly distributed, some MPI ranks may be inactive because the Master is waiting for other subdomains to be completed (i.e., refined).
- Increasing the background mesh's resolution and the octree's depth results in a larger number of subdomains, which can potentially increase concurrency. However, this also means less refinement work is left for the Workers. In addition, there is a clear tradeoff between the concurrency level and the data movement to computation ratio. As concurrency increases, the ratio also rises, ultimately leading to communication costs outweighing the costs of parallel computing. This can be seen from strong speedup analysis and might not be apparent in a weak scalability study performed in [12].

3. **Performance Evaluation**

At Old Dominion University, we have access to two HPC clusters used to evaluate the MPI-MW and PODM codes. The first one, Turing: is a 250-node cluster consisting of Intel(R) Xeon(R) (E5-2660, E5-2660 v2, E5-2670 v2, E5-2698 v3, E5-2683 v4) 128 GB CPUs ranging between 16 to 32 cores spread among two sockets (2 NUMA nodes). The second platform, Wahab: is a 200-node cluster that utilizes Intel(R) Xeon(R) Gold 6148 @ 2.4 GHz CPUs of 40 cores each in two sockets (4 NUMA nodes). In [12], we utilized the Turing cluster but opted for the new Wahab machine for this study. For all our experiments, the input image used is the abdominal atlas CT obtained from IRCAD Laparoscopic Center [35]. Figure 1 shows 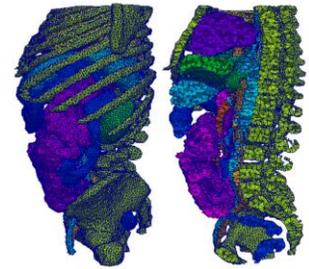

*Figure 1. An example mesh was created with the abdominal atlas image as input.*

an example of the mesh created using this input. The input images are decomposed using the same uniform octree as in [12]. We conducted experiments using two different data decompositions for an octree of depths 3 and 4, resulting in 512 ($8^3$) and 4096 ($8^4$) subdomains, respectively. The results are consistent with the finding in [12] and are briefly summarized below.

First, we check the correctness and quality of distributed memory MPI-MW PDR.PODM implementation against the shared memory PODM code presented in [15]. There are many ways to measure the quality of the mesh; in this work with the quality of the mesh, we refer to the quality of each element in the mesh which is measured by the dihedral angle distribution. Figure 2 depicts the dihedral angle distribution for PDR.PODM (left) and PODM (right); both histograms are almost identical, as expected. In contrast to data reported in [12], some slivers (dihedral angles close to 0° or 180°) are found in the mesh; they constitute less than 0.001% of the total –in this study, we did not attempt to remove slivers by adjusting the parameters of PODM– to the best of our knowledge, still there are no theoretical guarantees to eliminate slivers. PODM uses a heuristic to minimize the number of slivers [15]; sometimes, the heuristic with the "right" choice of parameters can return tetrahedra with a lower bound at 2º [12].

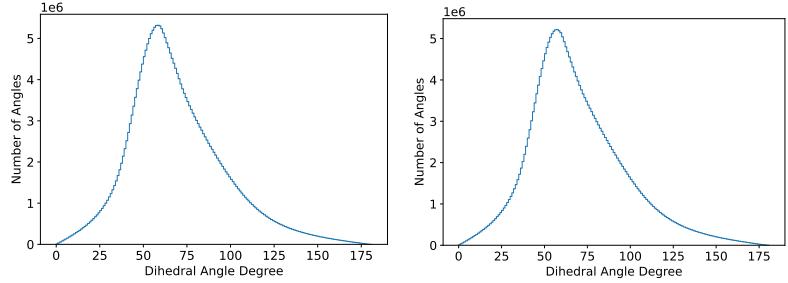

Figure 2. (Left) Dihedral angle distribution of the final mesh created by the PDR.PODM distributed memory implementation on several nodes. (Right) Dihedral angle distribution of the final mesh created by the PODM on a single node. The final mesh's total size is approximately 47 million elements.

In the remaining section, we will analyze the overheads of the parallel MPI-MW PDR.PODM implementation and strong scalability metric. In this study, we prefer to use strong scalability because in a traditional software pipeline with a field solver [36], a given fixed-size mesh is refined to meet certain error complexity, and the final mesh size depends on the application and error metrics instead of the number of cores used by the solver and/or end-to-end application.

For the strong scaling studies, we kept the size of the mesh generated constant (approximately 47 million elements). Both overheads and strong scalability are studied for single- and multi-node platforms with core numbers varying from 10 to 800. The pre-processing and parallel meshing time is negligible compared to distributed memory overheads (e.g., communication and idle times). When dealing with larger mesh sizes, the percentage of MPI-MW PDR.PODM will continue to follow the same pattern as before. This indicates that they will still take up much execution time and be much larger than the 50% reported in [12].

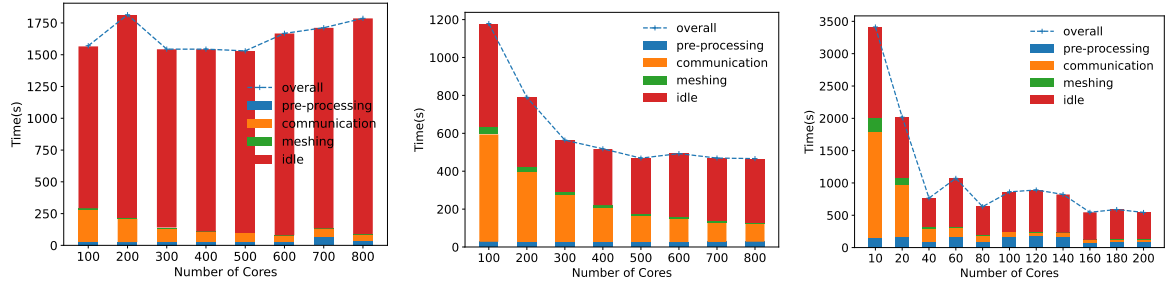

Figure 3. Breakdown of the running time for d=3 (left) and d=4 (center), where each MPI rank utilizes ten threads. Breakdown of the running time for d=4, where each MPI rank utilizes one thread (right). The final mesh consists of ≈ 47mil elements for all runs.

Figure 3 depicts the performance breakdown for the two different domain decomposition configurations using the same method (octree-based decomposition described in [12]) but varying the number of subdomains. The breakdown of the running time consists of four parts: (i) the pre-processing time is the time that the master process (i.e., host node) spends on loading an image from disk, constructing an octree, creating the coarse background mesh, assigning the elements of the coarse mesh to subdomains (partitioning); (ii) the meshing time is the time that a process spends on mesh refinement (more precisely, the PODM time of an MPI process);

(iii) the communication time is the time that a process spends on task requests and data movement, i.e., packing and unpacking data; (iv) the idle time is the time that a process waits in the waiting list and does not perform any mesh refinement work. Each bar is the sum of the time a process spends on each part for all iterations of PDR.PODM algorithm[3] will continue refinement until termination, which means there should be no more elements to refine. The process (worker) requests a subregion in each iteration and refines the submesh inside the subdomain. We calculate the average time of each part for all processes. Other metrics like min and max execution time could be used [12]; in this study, we are interested in whether the MPI-MW PDR.PODM overheads justify the further development of the method, and thus the average analysis will suffice.

It is important to note that there are two key parameters at the parallel mesh generation level that can improve parallel efficiency: (i) surface/volume ratio and (ii) concurrency, i.e., the number of subdomains that could be refined independently to each other —a small surface/volume ratio for a large number of independent subdomains helps to amortize the parallel computing costs due to communication and data dependencies of the PDR algorithm. On the left side of Figure 3, it is evident that low subdomain numbers (approximately 500) result in idle time (in red) dominating the execution time, owing to insufficient concurrency at the PDR level. When we move to the center of Figure 3, with eight times more subdomains, resulting in a total of around 4000, communication time (in orange) increases substantially (in contrast to d = 3) as expected (due to a larger surface/volume ratio). Nonetheless, idle time remains present and significant, even though we utilize 10 threads per MPI rank to minimize these overheads.

It appears that the distributed memory PDR.PODM algorithm is constrained by data dependencies[3], which continue to restrict concurrency even with an 8x increase in over-decomposition. For 100 cores, about 50% of the overheads in Figure 3 (center) are caused by idle time due to a lack of concurrency. At the same time, the shared memory implementation of PODM can explore quite effectively up to 144 cores [15]. Despite the large idle and communication overheads, even in the case of d = 4, some relative improvement (about 2x) in the performance of PDR.PODM is observed (i.e., the execution time for 100 vs. 500 cores).

Figure 3 (right) displays the execution time of an experiment where each MPI rank has only one thread (in contrast to the ten threads shown in Figure 3 left and center). The results show that in-node parallelism has a minimal effect when there are approximately 40 MPI ranks. When there are around 100 ranks, the performance is similar between 1 and 10 threads per rank, with a difference of no more than 20%. This result is influenced not only by application-level parameters but also by other factors. One such factor is maintaining a lower thread ratio per rank, which allows more workers to execute independent subdomains and distribute communication overheads more evenly. Additionally, it helps the scheduler allocate more ranks on the same or nearby hardware nodes, reducing communication overheads even further. By exploring parallelism solely on the rank level, thread interference (as discussed in Section 3.2) could be minimized and eliminated in the case of one thread per rank.

---

[3] For the detailed description of the PDR.PODM iterative refinement algorithm see [12] and [29].

Based on the data presented in Figure 3, we have observed that the optimal performance is achieved with an octree of level four, which consists of 4096 subdomains. Therefore, unless otherwise specified, the octree depth for all subsequent experiments will be set to four. This finding is consistent with previous research cited in [12].

In Figure 4, we provide a detailed analysis of the overheads, highlighting the trade-off between polling time and waiting-for-work time at the MW-MPI level. It's important to note that for the 100-core scenario (10 ranks x 10 threads), the polling time dominates because workers cannot respond to data requests from other workers while they are refining and packing/unpacking subdomains. As the number of ranks increases (to 500 and 800), a worker's total time spent in these processes is reduced since each worker will be responsible for a significantly smaller number of subdomains. Moreover, due to the balanced distribution of work (along with the respective data) to workers, the burden of serving data to other workers is distributed as more workers are included in the execution. As a result of these two aspects, the overall time spent in polling is reduced substantially when the number of workers increases. On the other hand, by increasing the number of workers (MPI ranks), the time spent waiting for a work assignment is increased due to the data dependencies among the available pieces of work. Since the size of the problem remains the same, the amount of work that can be executed concurrently reaches the point of saturation (at about 500 MPI ranks) before all the available ranks are utilized – clearly shown in the case of 800 MPI ranks, resulting in an underutilization of the system with a significant number of workers waiting for a work assignment. To summarize the data in Figure 4, there is a sweet spot for the number of workers needed for a fixed problem (mesh size), regardless of the number of subdomains. Going beyond this point may not improve performance.

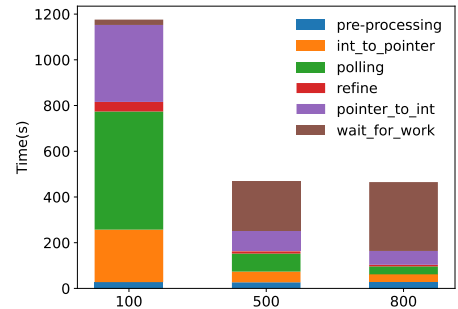

Figure 4. A detailed breakdown of the execution time

### 3.1 Single-Node Vs. Multi-Node Performance Evaluation

Previously, we observed that certain application/software stack parameter values can improve performance. But it's still unclear how this improvement compares to the performance of single-node multi-core PODM code presented in [15]. In this section, we evaluate the performance of two different memory models: distributed memory using the MPI-MW PDR.PODM described in [12] and shared memory PODM described in [15] run on a single multi-core node.

**Single-Node Evaluation.** The shared memory PODM takes advantage of multi-threading to explore concurrency, which allows for direct memory access and results in super-linear speedup for up to 16 cores, linear speedup for up to 64 cores, and parallel efficiency of 0.93 for up to 128 cores [15]. On the other hand, PDR.PODM utilizes MPI and multi-threading to explore concurrency at coarse and fine-grain levels by utilizing: (i) gather/scatter operations and (ii) introduces global indexing to manage data movement between subdomains (i.e., coarse-level) and PODM at the fine-level.

In a previous study [15], we found that hyper-threading can improve the use of shared memory resources such as the TLB, LLC, and pipeline, but performance slows down significantly after reaching 64 cores. To address this issue for systems with more than 64 cores, we conducted another study [29] where we implemented explicit data management through a 2-level decomposition approach --since it theoretically could improve the TLB and LLC overheads.

Although the data in Figure 3 and [29] show promise on their own, the results from Table 1 raise an important question: does the MPI-MW PDR.PODM implementation with explicit data management meet expectations for a fixed size problem? Our observation from running the MPI-MW PDR.PODM and PODM on a single multi-core node is that the additional overheads (i) and (ii) are significant, resulting in a 4x to 7x slowdown, as shown in Table 1.

Table 1. Execution time for PODM and PDR.PODM on a single multi-core node. Both runs use the same input and generate approximately the same mesh, about 47 million elements.

|  | Cores | No. of MPI Ranks | Execution Time (sec) | Platform |
|---|---|---|---|---|
| PODM | 1 | 0 | 2105.6 | Shared Memory |
| PODM | 40 | 0 | 90.3 | Shared Memory |
| PDR.PODM | 40 | 40 | 666.3 | Distributed Memory |
| PDR.PODM | 40 | 20 | 370.0 | Distributed Memory |

Table 2. Execution time for PDR.PODM on potentially multiple nodes as the cluster scheduler allocates the cores. All runs use the same input and generate approximately the same mesh, about 47 million elements.

| Cores | Ranks – Threads | Execution Time (sec) |
|---|---|---|
| 10 | 10 – 1 | 1569.2 |
| 10 | 5 – 2 | 1605.5 |
| 20 | 20 - 1 | 873.9 |
| 20 | 10 – 2 | 797.7 |
| 20 | 5 - 4 | 2347.7 |
| 40 | 40 – 1 | 679.3 |
| 40 | 20 – 2 | 794.1 |
| 40 | 10 – 4 | 1312.3 |
| 40 | 5 – 8 | 4435.1 |

**Multi-node Evaluation.** After discovering that the MPI-MW PDR.PODM costs are significantly higher than the communication costs for PODM on a single DSM node, the following question was raised: why wait or pay the high price for a single high-core (DSM) node when multiple multi-core nodes are more accessible and perhaps more cost-effective? However, our findings from Table 2 suggest that multiple nodes can make the situation even more challenging. We conducted tests with a fixed number of 40 cores and let the cluster scheduler assign MPI-rank cores per node according to current utilization. The results were inconsistent, but we noticed that allocating more than two cores per rank (i.e., PODM process) led to a deterioration in overall performance. This could be due to the cluster scheduler or the

PDR.PODM overheads are becoming more dominant as the MPI implementation no longer relies on optimized shared memory access.
In summary, when performing on multiple nodes, the overall performance of MPI-MW PDR.PODM is approximately 7 times slower than PODM's performance on a single node with the same number of cores and a fixed size problem.

**3.2 Thread Interference between Communication and Meshing Routines**

Earlier, we have seen the interplay between the polling and wait-for-work time and how they can be optimized by choosing "optimum" values for the application/software/hardware stack. However, for an optimum over-decomposition (and thus surface/volume ratio), one could try "hide" or tolerate some of the communication costs by using the available multi-threading in multi-core nodes. To tolerate some of the communication overheads, we modified the communication-related routines of the original PDR.PODM implementation is presented in [12] by introducing multi-threading in the packing/unpacking routines. *However, given that PODM also uses multi-threading, there is a potential for thread interference between the communication and meshing tasks.* It is important to note here that the number of threads used for packing/unpacking is equal to the number of available cores per rank and does not consider that PODM's refining threads are also present. Thus, potentially, the number of active threads at a time could be double the number of the available cores in a rank, leading to core oversubscription effects. However, as the performance data in Table 3 show, using the sequential version (as was the case in [12]) of these routines achieves significantly worse results.

*Table 3. Execution time of the fixed-size problem (47M elements) using threaded and non-threaded packing and unpacking routines for different Ranks-Cores configurations.*

| Cores | Ranks – Threads | Execution Time with mpirun (sec) | Execution Time (sec) for nonthreaded-(un)packing |
|---|---|---|---|
| 60 | 30 – 2 | 370.0 | 773.5 |
| | 15 - 4 | 450.1 | 1124.1 |
| 80 | 40 – 2 | 483.5 | 700.6 |
| | 20 – 4 | 686.9 | 1049.1 |
| 100 | 50 – 2 | 333.4 | 637.4 |
| | **25 - 4** | **357.8** | **863.9** |
| 120 | 60 – 2 | 418.8 | 606.2 |
| | **30 – 4** | **488.6** | **843.2** |
| 140 | 70 – 2 | 326.0 | 626.7 |
| | **35 – 4** | **343.4** | **734.5** |
| 160 | 80 - 2 | 408.5 | 639.7 |
| | **40 - 4** | **473.7** | **774.26** |
| 180 | 90 - 2 | 288.6 | 572.2 |
| | **45 - 4** | **317.1** | **696.5** |
| 200 | 100 - 2 | 353.4 | 560.4 |
| | **50 - 4** | **409.9** | 801.0 |

In this section, we evaluate the MPI-MW PDR.PODM algorithm/implementation performance again, but with the single-threaded versions for the packing/unpacking routines, i.e., there is no interference between communication threads and PODM's fine-grained threading.

Table 3 shows execution times for multiple nodes using 2-core and 4-core MPI ranks. Column 3 reports data on multi-threading to improve communication, while column 4 reports data from the original non-threading code for packing and unpacking from [12]. A more comprehensive data set is presented in Table 4 (Appendix I, where for additional validation SRUN is used). This table demonstrates the algorithm's performance under three different execution modes: (1) multi-threaded meshing and communication routines using mpirun[4], and no-threading for communication. Many important points can be derived from this quite comprehensive table; we underline two:
- Restricting packing/unpacking routines to only utilize a single thread significantly impacts the overall performance, which is to be expected since the cost for those two operations is the dominating term of execution time.
- We still observe counterintuitive results, with the overall time increasing when keeping the number of MPI ranks constant and adding more threads (for refinement-only) per rank, even though there is no case of thread interference. This might be due to the scheduler assigning MPI ranks further away in the hardware topology; by requesting more threads per MPI rank, the scheduler can no longer "fit" as many ranks on the same or neighboring hardware nodes, thus, the overheads of inter-node communication increase due to the overall traffic from other uses in the cluster.

Using sequential refinement but parallel packing/unpacking would be another interesting experiment. However, because the parallel implementation of these routines depends on structures only exposed when parallel refinement is used, it could not be investigated. Utilizing a tasking approach with a shared pool of threads for refinement and packing/unpacking is the best implementation approach but is outside the scope of this work and will be explored elsewhere. These issues must be studied further as we move HPC applications from dedicated computing facilities to Cloud platforms.

**4. Discussion**

This study aims to understand the nature of the MPI-MW overheads and potential scalability for PDR.PODM uses a strong speedup metric (i.e., fixed size problem) as opposed to a weak scalability study (i.e., increase the mesh size along with the number of cores, but keep the per core mesh size fixed) presented in [12]. We summarize our experience with several remarks:

**Remark 1.** Over-decomposition helps the MPI-MW PDR.PODM distributed memory implementation to a degree, but a careful balance needs to be considered. In this study, we observe about 2x relative improvement[5] in execution time by comparing the results from 500 cores for runs with 512 subdomains (one per core) and 4096 subdomains (8 per core).

---

[4] MPIRUN is a command that controls aspects of MPI program execution. Another similar control command is SRUN.
[5] Relative improvement (or speed up) is defined by comparing the same implementation for two different core counts as opposed to the (absolute) improvement (or speed up) where a comparison is made against the best-known parallel implementation, in this case PODM [13].

**Remark 2.** We see the fundamental tradeoffs in parallel computing between granularity and concurrency: one could decrease the granularity (at the domain decomposition level) for a potential increase in the concurrency, but this will introduce more overheads in terms of communication. If there is not enough work (number of subdomains to be refined in parallel) compared to the number of processing units (hardware cores) to exploit the available concurrency in the system, a data decomposition with smaller granularity (i.e., a larger number of subdomains) does not improve the algorithm's performance. This is very clear from Figure 3 (center) and for core numbers from 500 to 800.

Based on these two observations, it's reasonable to ask if either explicit data management using MPI or utilizing multiple nodes to decrease execution time is worth it, even if the application's memory usage isn't significant enough to warrant it. The following two remarks sum up our experience:

**Remark 3.** While explicit data management in the distributed memory MPI-MW implementation theoretically can reduce TLB and LLC overheads in PODM, certain drawbacks still come with it. These include idle time due to limited concurrency caused by data dependencies between data-decomposed work units (subdomains) in PDR and communication-related overheads resulting from the cost of gather/scatter operations and global indexing for managing data movement between subdomains.

**Remark 4.** When using multiple nodes, the distributed memory MPI-MW implementation experiences a 2x slowdown compared to using the same number of cores on a single node. This is evident in the "best" entries 20-2 (see Tables 1 and 2). It's important to note that the shared memory PODM implementation is several times (about 7x) faster than the MPI-MW PDR.PODM implementation using multiple nodes.

To tolerate communication latencies, multi-threading helps (see Table 3), but thread interference is a real issue, and our experience is summarized below:

**Remark 5.** The number of threads for communication and mesh refinement varies at runtime due to the degree (connectivity) of subdomains and refinement method. Lower and upper bounds could be estimated, but generally, it is difficult to control the oversubscription of threads from these modules given a fixed number of hardware cores –determined at the initialization of the parallel program. So, when selecting the best hybrid algorithm configuration, it's important to weigh the trade-offs between thread interference at the mesh refinement module and communication-related modules.

This study highlights the possibility of coming to incorrect conclusions if one looks at a relative (i.e., weak scalability analysis), and thus, the data presented in [12] should be carefully evaluated and examined in the context of this experience, too. While the performance of using 10 to 40 (and similarly from 100 to 500) cores has shown some relative improvement (see Figure 3), it is important to note that this study (using strong speed-up analysis) demonstrates the MPI-MW implementation and the PDR.PODM presented in [12] is more than twice (at best) slower than the PODM approach presented [15]. This suggests that PDR.PODM framework and its MPI-MW implementation may not reliably scale the PODM mesh generation and refinement method if the goal is to reduce execution time. If memory footprint is the issue, alternative options like Out-of-Core (OoC) presented in [37, 38] must be considered.

Our objective of implementing distributed memory codes using MPI-MW approach and highly efficient 3D shared memory implementations like PODM, which are not designed for multiple nodes, was not achieved. In addition, labor may be reduced, but at a high cost. Overheads for preparing and moving data account for more than 50% of execution time on multi-node platforms. Communication costs and idle time due to data dependencies in PDR.PODM framework make the distributed version over 7x slower than shared memory code on the same number of cores, requiring 8x more resources and electricity to improve end-to-end execution time.

## 5. Conclusions and Future Work

Given the experience from the specific implementation (presented in [12]) of PODM on distributed memory machines, we must question whether: (i) scalable (weak) speedup studies (i.e., increase the problem size as we increase the number of cores) provide any (useful) insight for distributed memory parallel mesh generation and by extension, (ii) the full scalability of PDR.PODM and more general parallel mesh generation is a worthwhile goal for parallel machines (with $10^6$ or more cores) in the exascale era, and if not, then (iii) one needs to ask: what is the alternative?

We conclude this paper with some final remarks regarding those questions. The scalability of parallel mesh generation methods (being data-intensive as opposed to field solvers, which are compute-intensive) depends on the ratio of data movement to computation for refinement. As the parallel mesh generation and (even more end-to-end function approximation) progresses, the opportunities for refinement are reduced after some point, and data movement is unnecessarily increased (especially with the PDR framework as we witnessed in this study for PDR.PODM) with minimal impact on the overall performance. In a scenario where the end goal is to approximate a fixed (unknown) function, the method (field solver) determines the size of the mesh required to resolve all the features of the function for a given tolerance (error). In this case, the question is, what is the optimum use of resources (cores & network) and electric power to minimize walk-clock time? This study (with the strong speedup analysis above and Figures 3 and 4) points to future directions regarding what is essential and needs to be optimized.

When dealing with the complex task of parallel mesh generation, allocating only a portion of the available resources for function approximation is crucial. Proper (strategic) mapping of the mesh generation problem to the available hardware for function approximation is key to ensuring success. Existing mapping methods, such as [39], can be useful. Minimizing the setup time for function approximation codes is best achieved using a strong scalability metric or equivalent since the problem is reduced to minimize the end-to-end walk-clock time for setting up (or refreshing) the data structure for the function approximation codes [38]. This leads us to wonder if this holds for other similar data-intensive unstructured parallel mesh generation algorithms and software. In our future work, we should provide more insight into parallel Advancing Front and metric-based anisotropic using local reconnection schemes [41] and [42].

In the future, also we plan to consider the parallelization of two existing sequential image-to-mesh conversion methods [32-34], using the lessons learned from this study. It is worth noting that most image-to-mesh conversion codes used in healthcare are primarily focused on real-time applications, such as surgery simulations and image-guided surgery [43-45] for 3D static images and 4D moving images [46]. These applications require a fixed, small-size problem to ensure the best results within the time constraints of the procedure.

**Acknowledgments.** We want to acknowledge Dr. Daming Feng and Dr. Andrey Chernikov for their work on an earlier version of the MPI-MW code (presented in [12]), which was used after several modifications required to address several issues to perform this study.

## 6. Appendix I

Table 4. Execution time of the fixed-size problem (47M elements) using threaded and non-threaded packing and unpacking routines for different Ranks-Cores configurations.

| Cores | Ranks – Threads | Execution Time with mpirun (sec) | Execution Time with srun (sec) | Execution Time (sec) for nonthreaded-(un)packing |
|---|---|---|---|---|
| 60 | 60 – 1 | 899.8 | 573.3 | 643.9 |
|    | **30 – 2** | **370.0** | **388.8** | **773.5** |
|    | 15 - 4 | 450.1 | 478.8 | 1124.1 |
| 80 | 80 – 1 | 551.1 | 532.9 | 623.2 |
|    | **40 – 2** | **483.5** | **857.1** | **700.6** |
|    | 20 – 4 | 686.9 | 919.3 | 1049.1 |
|    | 10 - 8 | 1374.9 | 1740.2 | 1991.0 |
| 100 | 100 – 1 | 690.6 | 542.8 | 641.6 |
|    | **50 – 2** | **333.4** | **357.3** | **637.4** |
|    | 25 - 4 | 357.8 | 378.2 | 863.9 |
| 120 | 120 – 1 | 719.3 | 569.1 | 617.2 |
|    | **60 – 2** | **418.8** | **702.4** | **606.2** |
|    | 30 – 4 | 488.6 | 717.6 | 843.2 |
|    | 15 – 8 | 923.6 | 1156.1 | 1430.4 |
| 140 | 140 – 1 | 647.2 | 520.9 | 572.5 |
|    | **70 – 2** | **326.0** | **319.2** | **626.7** |
|    | 35 – 4 | 343.4 | 355.6 | 734.5 |
| 160 | 160 - 1 | 462.9 | 488.6 | 577.8 |
|    | **80 - 2** | **408.5** | **319.2** | **639.7** |
|    | 40 - 4 | 473.7 | 655.7 | 774.26 |
|    | 20 - 8 | 745.9 | 919.7 | 1163.7 |
| 180 | 180 - 1 | 502.6 | 476.4 | 574.3 |
|    | **90 - 2** | **288.6** | **313.4** | **572.2** |
|    | 45 - 4 | 317.1 | 342.1 | 696.5 |
| 200 | 200 - 1 | 457.4 | 486.2 | 533.1 |
|    | **100 - 2** | **353.4** | **635.3** | **560.4** |
|    | 50 - 4 | 409.9 | 601.8 | 801.0 |
|    | 25 - 8 | 618.9 | 773.1 | 975.7 |

# References

[1] Houstis, Elias N.; Rice, John R.; Chrisochoides, N. P.; Karathansis, H. C.; Papachiou, P. N.; Vavalis, E. A.; and Wang, K., "Parallel (//) Ellpack PDE Solving System" (1989). Department of Computer Science Technical Reports. Paper 778. https://docs.lib.purdue.edu/cstech/778

[2] Chrisochoides, N. P.; Houstis, C. E.; Houstis, Elias N.; Papachiou, P. N.; and Kortesis, S. K., "DOMAIN DECOMPOSER: A Software Tool for Mapping PDE Computations to Parallel Architectures" (1990). Department of Computer Science Technical Reports. Paper 27. https://docs.lib.purdue.edu/cstech/27

[3] Nikos Chrisochoides. 1992. On the mapping of partial differential equation computations onto distributed memory MIMD parallel machines. Ph.D. Thesis, Purdue University, W/ Lafayette, IN, (August 1992).

[4] Nikos Chrisochoides, Elias Houstis, and John Rice. 1994. Mapping algorithms and software environment for data parallel PDE iterative solvers. J. Parallel Distrib. Comput. 21, 1 (April 1994), 75–95. https://doi.org/10.1006/jpdc.1994.1043

[5] Nikos Chrisochoides, "An alternative to data mapping for parallel PDE solvers: parallel grid generation," Proceedings of Scalable Parallel Libraries Conference, Mississippi State, MS, USA, 1993, pp. 36-44, doi: 10.1109/SPLC.1993.365584.

[6] Nikos Chrisochoides. 2002. A new approach to parallel mesh generation and partitioning problems. Computational Science, Mathematics, and software. Purdue University Press, USA, (January 2002) 335–359, 2002.

[7] Kevin Barker, Nikos Chrisochoides, Jeff Dobbelaere, Demian Nave, and Keshav Pingali (2002), Data movement and control substrate for parallel adaptive applications. Concurrency Computat.: Pract. Exper., 14: 77-101. https://doi.org/10.1002/cpe.617

[8] Nikos Chrisochoides, Kevin Barker, Demian Nave, Chris Hawblitze. 2000. Mobile object layer: A runtime substrate for parallel adaptive and irregular computations Advances in Engineering Software 31 (8-9), 621-637

[9] Kevin Barker, Nikos Chrisochoides. 2003. An evaluation of a framework for the dynamic load balancing of highly adaptive and irregular parallel applications. Proceedings of the 2003 ACM/IEEE Conference on Supercomputing, 45

[10] Polykarpos Thomadakis, Christos Tsolakis, Nikos Chrisochoides. 2022. Multithreaded runtime framework for parallel and adaptive applications Engineering with Computers 38 (5), 4675-4695

[11] Polykarpos Thomadakis, Nikos Chrisochoides. 2023. Toward runtime support for unstructured and dynamic exascale-era applications. The Journal of Supercomputing, 1-28

[12] Daming Feng, Andrey Chernikov, Nikos Chrisochoides. 2018. A scalable hybrid parallel Delaunay image-to-mesh conversion algorithm for distributed-memory clusters. Computer-Aided Design 103, 34-46

[13] Nikos Chrisochoides. 2006. Parallel mesh generation. Numerical solution of partial differential equations on parallel computers, Vol. 51, pages 237-264.

[14] Nikos Chrisochoides and Demian Nave. Parallel Delaunay mesh generation kernel. International Journal for Numerical Methods in Engineering 58 (2), 161-176

[15] Panagiotis Foteinos and Nikos Chrisochoides. 2014. High-quality real-time Image-to-Mesh conversion for finite element simulations. Journal of Parallel and Distributed Computing 74 (2), (February 2014), 2123-2140, https://doi.org/10.1016/j.jpdc.2013.11.002

[16] Andrey Chernikov, Nikos Chrisochoides. Practical and efficient point insertion scheduling method for parallel guaranteed quality Delaunay refinement. Proceedings of the 18th annual international conference on Supercomputing, 48-57, 2004.

[17] Nikos Chrisochoides, Andrey Chernikov, Thomas Kennedy, Christos Tsolakis, and Kevin Garner. Parallel Data Refinement Layer of a Telescopic Approach for Extreme-scale Parallel Mesh Generation for CFD Applications. Published in AIAA Aviation Forum 2018, Atlanta, Georgia, June 2018

[18] Andrey Chernikov and Nikos Chrisochoides. Algorithm 872: Parallel 2D Constrained Delaunay Mesh Generation. Published in ACM Transactions on Mathematical Software, Volume 34, No. 1, pages 6 -- 25, January 2008

[19] Christos Tsolakis, Andrey Chernikov and Nikos Chrisochoides. Parallel Constrained Delaunay Meshing Algorithm in Three Dimensions. Published in 2017 Modeling, Simulation, and Visualization Student Capstone Conference, Suffolk, VA, April 2017.

[20] Leonidas Linardakis, Nikos Chrisochoides. Graded Delaunay decoupling method for parallel guaranteed quality planar mesh generation. SIAM Journal on Scientific Computing 30 (4), 1875-1891

[21] Maria Rivara, C Calderon, A Fedorov, N Chrisochoides. Parallel decoupled terminal-edge bisection method for 3D mesh generation. Engineering with Computers 22, 111-119

[22] Andrey Chernikov and Nikos Chrisochoides. Three-dimensional semi-generalized point placement method for Delaunay mesh refinement. Proceedings of the 16th International Meshing Roundtable, 25-44

[23] Andrey Chernikov and Nikos Chrisochoides. Parallel guaranteed quality Delaunay uniform mesh refinement. SIAM Journal on Scientific Computing 28 (5), 1907-1926

[24] Nikos Chrisochoides and Florian Sukup. Task Parallel Implementation of the BOWYER-WATSON Algorithm. In Proceedings of 5th Intern. Conference on Numerical Grid Generation in Comput. Fluid Dynamics and Related Fields, pp 773--782, 1996.

[25] Christos Tsolakis, Polykarpos Thomadakis, Nikos Chrisochoides. Tasking framework for adaptive speculative parallel mesh generation. The Journal of Supercomputing 78 (5), 1-32.

[26] Nikos Chrisochoides. Multithreaded Model for Load Balancing Parallel Adaptive Computations on Multicomputers. In Journal of Applied Numerical Mathematics Vol. 20, pp 349--365, April 1996.

[27] Polykarpos Thomadakis, Christos Tsolakis and Nikos Chrisochoides. Multithreaded Runtime Framework for Parallel and Adaptive Applications. Published in Engineering with Computers, Publisher Springer, 2022

[28] Daming Feng, Andrey Chernikov, Nikos Chrisochoides. 2016. Two-level locality-aware parallel Delaunay image-to-mesh conversion. Parallel Computing 59, 60-70

[29] Daming Feng, Christos Tsolakis, Andrey Chernikov, Nikos Chrisochoides. 2017. Scalable 3D hybrid parallel Delaunay image-to-mesh conversion algorithm for distributed shared memory architectures. Computer-Aided Design 85, 10-19

[30] Andrey Chernikov and Nikos Chrisochoides. Three-Dimensional Delaunay Refinement for Multi-Core Processors. Published in 22nd ACM International Conference on Supercomputing, pages 214 -- 224, Island of Kos, Greece, June 2008

[31] Andrey Chernikov, Nikos Chrisochoides. 2012. Generalized insertion region guides for Delaunay mesh refinement SIAM Journal on Scientific Computing 34 (3), A1333-A1350

[32] AN Chernikov, NP Chrisochoides. Multitissue tetrahedral image-to-mesh conversion with guaranteed quality and fidelity. SIAM Journal on Scientific Computing 33 (6), 3491-3508

[33] Y Liu, P Foteinos, A Chernikov, N Chrisochoides. Mesh deformation-based multi-tissue mesh generation for brain images. Engineering with Computers 28 (4), 305-318

[34] Fotios Drakopoulos, Ricardo Ortiz, Andinet Enquobahrie, Deanna Sasaki-Adams and Nikos Chrisochoides. Tetrahedral Image-To-Mesh Conversion Software for Anatomic Modeling of Arteriovenous Malformations. Published in 24th International Meshing Roundtable, October 2015

[35] Ircad Laparoscopic Center, http://www.ircad.fr/softwares/3Dircadb/3Dircadb2 (2013)

[36] [Christos Tsolakis, Nikos Chrisochoides. 2019. Anisotropic mesh adaptation pipeline for the 3d laminar flow over a delta wing. SpringSim-ANSS, (April 2019), Suffolk, VA.

[37] Andriy Kot, Andrey Chernikov, Nikos Chrisochoides. (2011). Effective out-of-core parallel Delaunay mesh refinement using off-the-shelf software. ACM Journal of Experimental Algorithmics. Vol 16, No.: 1.5 pp 1.1–1.22, https://doi.org/10.1145/1963190.2019580

[38] A. Kot and N. Chrisochoides, "Green" multi-layered "smart" memory management system," Second IEEE International Workshop on Intelligent Data Acquisition and Advanced Computing Systems: Technology and Applications, 2003. Proceedings, Lviv, Ukraine, 2003, pp. 372-377, doi: 10.1109/IDAACS.2003.1249589.

[39] Nikos Chrisochoides, Elias Houstis, Cathryn Houstis. Geometry-based mapping strategies for PDE computations. Proceedings of the 5th international conference on Supercomputing, 115-127, 1991.

[40] Nikos Chrisochoides, Cathryn Houstis, Elias Houstis, Panos Papachiou, Sotiris Kortesis, and John Rice. Domain Decomposer: A Software Tool for Partitioning and Allocation of PDE Computations Based on Geometry Decomposition Strategies. In Proceedings of the 4th International Symposium on Domain Decomposition Methods, Moscow, USSR, pp 341--357, SIAM Publications, 1991.

[41] Fotis Drakopoulos, Christos Tsolakis, Nikos Chrisochoides. 2019. The fine-grained speculative topological transformation scheme for local reconnection methods. AIAA Journal 57 (9), (July 2018), 4007-4018

[42] Christos Tsolakis, Nikos Chrisochoides, Mike Park, Andinet Loseille, Todd Michal. Parallel anisotropic unstructured grid adaptation. AIAA Journal 59 (11), 4764-4776

[43] Mike Audette, Andrey Chernikov, Nikos Chrisochoides (2012). A review of mesh generation for medical simulators. In Handbook of Real-World Applications in Modeling and Simulation edited by John A. Sokolowski, Catherine M. Banks

[44] Liu Yixun, Kot Andriy, Drakopoulos Fotis, Yao Chengjun, Fedorov Andrey, Enquobahrie Andinet, Clatz Olivier, Chrisochoides Nikos. (2014). An ITK implementation of a physics-based non-rigid registration method for brain deformation in image-guided neurosurgery. Frontiers in Neuroinformatics 8, 33. DOI=10.3389/fninf.2014.00033

[45] F Drakopoulos, P Foteinos, Y Liu, NP Chrisochoides. (2014). Toward a real-time multi-tissue Adaptive Physics-Based Non-Rigid Registration framework for brain tumor resection. Frontiers in Neuroinformatics 8, 11. DOI=10.3389/fninf.2014.00011

[46] Panagiotis Foteinos, Nikos Chrisochoides. 2015. 4d space-time Delaunay meshing for medical images. Engineering with Computers 31, 499-511.